\documentclass[%
 reprint,
 amsmath,amssymb,
 aps,
]{revtex4-2}

\usepackage{graphicx}% Include figure files
\usepackage{dcolumn}% Align table columns on decimal point
\usepackage{bm,xcolor}% bold math
\usepackage{amssymb}
\usepackage{amsthm,amsmath}
\usepackage{bbm,hyperref}
\usepackage{braket,mathrsfs}
\usepackage[varg]{txfonts} 

\DeclareMathOperator{\Tr}{Tr}

\begin{document}

\preprint{APS/123-QED}

\title{Theory-independent realism}

\author{D. M. Fucci}
 \email{danilo@fisica.ufpr.br}
\author{R. M. Angelo}%
 \email{renato@fisica.ufpr.br}
\affiliation{%
 Department of Physics, Federal University of Paraná, P.O. Box 19044, 81531-980 Curitiba, Paraná, Brazil
}%

\date{\today}% It is always \today, today,

\begin{abstract}
The distinctive features of quantum mechanics, which set it apart from other physical theories, challenge our notions of realism. Recovering realism from purely philosophical grounds, a quantitative and operational criterion was proposed in the past, but solely for the context of quantum mechanics. We use a framework of generalized probabilistic theories to expand the notion of realism for a theory-independent context, providing a criterion uniquely based on the probabilities assigned to measurement outcomes. More so, using robustness and the Kullback-Leibler divergence, we propose quantifiers for the realism of arbitrary physical properties given a particular state of a generic physical theory. These theory-independent quantifiers are then employed in quantum mechanics and we investigate their relation with another well-established irrealism measure.
\end{abstract}

\maketitle

%\tableofcontents

\section{\label{sec:level1}Introduction}

Scientific knowledge is assumed to go beyond mere appearances and to convey reliable information about that which our senses cannot directly apprehend. In Physics, a large fraction of our scientific knowledge stands upon quantum theory, which remains coherent with the data acquired for more than a century. However, unlike classical mechanics, whose mathematical formulation aligns with our everyday experiences, quantum theory is challenging to make sense of. To abstract a consistent story from its formalism, or to interpret it, we are required to give up at least one of the metaphysical assumptions founding our classical intuition. One of such assumptions is realism.

Broadly speaking, realism tells us about the definiteness of the physical properties of a system at any instant of time independently of observers. Its seminal definition was given by Einstein, Podolsky, and Rosen (EPR) in 1935~\cite{Einstein1935}, without which the first hint of entanglement appearing in the literature would not have been possible. There, it refers to the capability of predicting with certainty the value assigned to a physical property without disturbing the system, such that, whenever this is the case, it is said that there is an element of reality assigned to that physical property. Conceiving an experiment where measurements are performed on two parts of an entangled system, with these measurements being events located outside of each other's light cones, EPR argue that incompatible observables should be simultaneous elements of reality. The authors then concluded that this would constitute a proof of the incompleteness of quantum mechanics. Put under scrutiny by Bell in 1964, this claim was discredited in Ref.~\cite{Bell1964}, where it was shown that the predictions made by quantum mechanics conflicted with any theory of local hidden variables. Quantum systems that violate Bell's inequalities, now understood to violate the hypothesis of local causality \cite{Bell2001}, have been experimentally verified through loophole-free tests~\cite{Hensen2015, Giustina2015, Shalm2015, Hensen2016, Rauch2018, Li2018}.

Even though it is possible to envision fully realist interpretations of quantum mechanics at the expense of other metaphysical requirements for the theory, such as in Bohmian mechanics~\cite{Bohm1952}, which sacrifices locality, the concept of realism underpins the debate around entanglement or Bell nonlocality. Later developments of the concept of realism, such as that proposed by Fine in Ref.~\cite{Fine1982}, refer to joint probability distributions assigned to the measurements of different observables. The probabilistic profile of the joint measurability of non-commuting observables contrasts with this definition, highlighting a clash between incompatibility and realism. Furthermore, the measurement problem, or what makes a measurement a measurement~\cite{Brukner2017}, is an issue that invites us to answer how realism, displayed in classical phenomena, emerges.

Newer proposals for realism criteria, such as the one devised by Bilobran and Angelo (BA) in Ref.~\cite{Bilobran2015}, make the concept more tangible. BA, in particular, do this by providing a quantifiable and operational criterion that was later employed in the development of a concept of nonlocality~\cite{Gomes2018, Gomes2019, Orthey2019, Fucci2019} and axiomatized~\cite{Orthey2022}. Further developments applied this notion to quantum resources theory~\cite{Costa2020} and foundational elements in quantum mechanics~\cite{Dieguez2018, Engelbert2020, Rudnicki2018, Savi2021, Engelbert2023,  Paiva2023, Mancino2018, Dieguez2022}. However, every proposal in the physics literature for a realism criterion pertains to the context of quantum mechanics. A definition for realism which is theory-independent is lacking, and this work aims at filling this gap.

Besides deepening our understanding of realism itself, a theory-independent approach allows us to better understand not only quantum mechanics \textit{per se} but also the relation between quantum mechanics and other physical theories. More so, it equips us with a framework suitable to scrutinize physical theories that are still to be developed, once quantum mechanics may not be the fundamental theory of Nature. Work \cite{Schmid2024} gives us a clear example of a theory-independent proposal shedding light on the concept of macrorealism~\cite{Legget1985}.

A concise and meaningful criterion for realism, fully encoded in a mathematical syllogism, that is independent of any particular theory but still pertains to the realm of physics, is introduced in this work through generalized probabilistic theories (GPT). The GPT framework is constructed as a mathematical structure that describes generic states, transformations, and measurements. In this framework, any physical theory can be viewed as a particular GPT defined by a set of rules that assign probabilities to outcomes for the measurements of physical properties. Our criterion relies on the probabilities themselves, rather than on the rules that determine them. Notwithstanding, we go further and, inspired by the works of BA, we propose two ways for quantifying, in a non-realist theory, the extent to which a state deviates from our criterion of realism. These sought-after measures are obtained using the concepts of robustness and the Kullback-Leibler divergence \cite{Kullback1951}, and they constitute two theory-independent quantifiers of irrealism.

The article is structured as follows. In Section~\ref{sec:level2} we review preliminary concepts, summarizing BA's criterion of realism and skimming through some building blocks of the GPT framework, presenting only the minimal amount necessary for the understanding of the following sections. Our criterion is then laid out in Section~\ref{sec:level3}, and in Section~\ref{sec:level4}, the theory-independent quantifiers of irrealism are presented along with some numerical analysis. They are tested in quantum mechanics and contrasted with BA's notion of irrealism. Section~\ref{sec:level5} closes the paper with our conclusions.

\section{\label{sec:level2}Preliminary Concepts}

\subsection{BA's criterion of realism}

Put forward in Ref.~\cite{Bilobran2015}, this is an operational criterion based on the premise that measurements are not able to change the epistemic description of an already installed reality. A protocol is devised in which a source prepares two sets of a very large number of identical copies of a bipartite quantum system, with one of the sets being intercepted by an agent performing an unrevealed measurement of an observable $A = \sum_a a A_a$, acting on partition $\mathcal{A}$, over each copy in the set. Here, $A_a$ represents the projectors of observable $A$ with assigned outcomes $a$. Quantum state tomography performed at the end will result in $\Phi_A(\rho)$ for the intercepted ensemble and $\rho$ for the non-intercepted one. If these two descriptions are indistinguishable, then $A$ is termed an element of reality. Mathematically, BA's criterion of realism is expressed as
\begin{equation} \label{eq:bilcrit}
    \rho = \Phi_A (\rho)
\end{equation}
with
\begin{equation} \label{eq: Phi_A}
    \Phi_A(\rho) \coloneqq \sum_a(A_a\otimes\mathbbm{1}_\mathcal{B})\,\rho\,(A_{a}\otimes\mathbbm{1}_\mathcal{B}) = \sum_{a}p_{a}A_{a} \otimes \rho_{\mathcal{B}|a}
\end{equation}
where $\mathcal{B}$ is the second partition of each system, $p_{a}$ is the probability associated to the outcome $a$, and $\rho_{\mathcal{B}|a}$ is the post-measurement state for the partition $\mathcal{B}$. A deviance measure from the criterion \eqref{eq:bilcrit} is then defined employing relative entropy:
\begin{equation} \label{eq: bil irr}
    \mathfrak{I}_A(\rho) \coloneqq S(\rho || \Phi_A(\rho)) = S(\Phi_A(\rho))-S(\rho),
\end{equation}
where $S(\rho)=-\Tr(\rho \log{\rho})$ stands for the von Neumann entropy of $\rho$ and $S(\rho||\varrho)=\Tr\big[\rho(\log{\rho}-\log{\varrho})\big]$ for the quantum relative entropy of $\rho$ and $\varrho$. The measure $\mathfrak{I}_A(\rho)$ reads as the irreality of $A$ given $\rho$. By now, irreality is a widely explored concept, supported by numerous investigations, both theoretical~\cite{Gomes2018, Gomes2019, Orthey2019, Fucci2019, Orthey2022, Costa2020, Dieguez2018, Engelbert2020, Rudnicki2018, Savi2021, Engelbert2023, Paiva2023} and experimental~\cite{Mancino2018, Dieguez2022}.

\subsection{Generalized probabilistic theories}

A general framework that has shown to be useful in dealing with generic physical theories, thus allowing a more solid background for talking about theory-independent properties, is called generalized probabilistic theories. It stands on three main building blocks: 

\begin{itemize}
\item[a)]{\it State spaces} generalize the concept of the set of all possible density operators in quantum mechanics. They may be understood as a set of equivalence classes of preparation procedures, a list of instructions to be followed wielding a particular physical state. Mathematically, it is expressed as a convex subset of a real finite-dimensional vector space.

\item[b)]{\it Effect algebras} map states inhabiting a state space into probabilities, generalizing what in quantum mechanics are the positive operator valued measures (POVMs). They are a set of equivalence classes of ``yes'' or ``no'' questions that may be made about a physical state in an experimental setting. So, let $\mathscr{K}$ be a state space and $E$ an effect algebra acting over $\mathscr{K}$. $E(\mathscr{K})$ is the set of all affine functions such that $f \colon \mathscr{K} \to [0,1]$.

\item[c)]{\it Channels} generalize quantum channels, transforming physical states from one system into physical states of another. For state spaces $\mathscr{K}_A$ and $\mathscr{K}_B$, a channel $\Phi$ is an affine map $\Phi \colon \mathscr{K}_A \to \mathscr{K}_B$.
\end{itemize}

Together with a composition rule, corresponding to the tensor product in quantum mechanics, which describes the formation of joint state spaces, those three elements give a mathematical framework capable of representing the operational features of arbitrary physical theories. A thorough introduction on the subject is given in Ref.~\cite{Plavala2023}.

\section{\label{sec:level3}Theory-Independent Realism}

In a theory where a given observable is real (has a definite value regardless of observation), the role of a measurement is merely to reveal the already established element of reality. If the information about the outcome of a performed measurement is lost or somehow inaccessible, then the state of knowledge remains the same as before the measurement. In other words, unrevealed measurements are innocuous in realist theories. This will be the fundamental premise behind the criterion we now propose.

To implement this principle, consider a generic state $\epsilon$ of a state space $\mathscr{K}$ with dimension $d$, and physical properties $\mathcal{X}=\{x_i\}_{i=1}^d$ and $\mathcal{Y}=\{y_j\}_{j=1}^d$ defined by the set of their possible outcomes. The criterion of realism requires that the probabilities assigned to any outcome $x_i$ upon the measurement of any property $\mathcal{X}$ on a state $\epsilon$ remains unchanged if an unrevealed measurement of $\mathcal{Y}$ was performed beforehand. Symbolically, if
\begin{equation} \label{eq:criterion}
    p_\epsilon(x_i) = p_{\Phi_\mathcal{Y}(\epsilon)}(x_i)\qquad \left(\forall\,\mathcal{X}=\{x_i\}_{i=1}^d\right),
\end{equation}
then $\mathcal{Y}$ is termed an element of reality given the state $\epsilon$. In the right-hand term, sub-index $\Phi_\mathcal{Y}(\epsilon)$ stands for a post unrevealed measurement of $\mathcal{Y}$ over state $\epsilon$. The term is expressed via conditional probabilities as
\begin{equation}\label{eq: Phi_Y(e)}
    p_{\Phi_\mathcal{Y}(\epsilon)}(x_i) \coloneqq \sum_{j} p_{\epsilon}(x_i|y_j) \,p_\epsilon(y_j).
\end{equation}
%
%Note the presence of a state $\tilde{\epsilon}$, generally different from $\epsilon$, in the conditional probability. This conceives cases where knowing a precise measurement outcome $y_j$ may alter the preparation $\epsilon$.

Note that the knowing of $y_j$ may alter the preparation $\epsilon$. The notation for conditional probabilities denotes that $\epsilon$ was the original state before the knowledge yielded by a precise measurement of $\mathcal{Y}$.
Criterion \eqref{eq:criterion} gives the core of this work. It does not demand every physical state $\epsilon$ to satisfy \eqref{eq:criterion}. When that is the case, that physical theory is considered realist. If only a set of states $\epsilon$ meets \eqref{eq:criterion}, those states are said to be $\mathcal{Y}$-realist.

Before moving on, it is opportune to note that our realism criterion can be validated through the assumption that joint probability distributions $p(x_i,y_j)$ exist for all $\mathcal{X}$, which makes contact with Fine's approach to determinism~\cite{Fine1982}. If $p(x_i, y_j)$ exists without any concern a priori with the order with which $\mathcal{X}$ and $\mathcal{Y}$ are probed, and if the definition of conditional probability holds, then one can write $p(x_i, y_j) = p_\epsilon(x_i|y_j) \,p_\epsilon(y_j)=p_\epsilon(y_j|x_i)\,p_\epsilon(x_i)$ (Bayes' rule). Since $\sum_j p_\epsilon(y_j|x_i) = 1$, substituting Bayes' rule into the definition \eqref{eq: Phi_Y(e)} immediately retrieves the realism criterion \eqref{eq:criterion}.

%We assume that knowing the outcome of an unrevealed measurement of a property $Y$ simply updates the probability distribution of the measured property without interfering with the state $\Phi_Y(\epsilon)$. The probability for getting an outcome $y_j$ in a revealed measurement protocol of $Y$ in a $Y$-realist state and $x_i$ afterwards reads
%\begin{equation} \label{eq: semijoint}
%    p_\epsilon (x_i ; y_j) = p_\epsilon (x_i | y_j) p_\epsilon (y_j).
%\end{equation}
%Here, $p_\epsilon (x_i ; y_j)$ is not meant to be read as the usual joint probability because, in this context, the order of the measurements matters. It is a joint probability conditioned to $X$ being measured after $Y$.

%For a state $\epsilon$ which is simultaneously a $Y$ and $Z$-realist state, revealed measurements of both properties allow for the coexistence of the identities \eqref{eq: semijoint} and
%\begin{equation}
%    p_\epsilon (x_i ; z_k) = p_\epsilon (x_i | z_k) p_\epsilon (z_k).
%\end{equation}
%Taking the particular case where $x_i$ is $y_j$ in the last equation retrieves the usual joint probability
%\begin{equation}
%    p_\epsilon (y_j , z_k) = p_\epsilon (y_j | z_k) p_\epsilon (z_k) = p_\epsilon (z_k | y_j) p_\epsilon (y_j)
%\end{equation}
%and also enables the validity of Bayes' theorem. Because for a realist theory, the joint probability is always valid, our criterion generalizes the idea proposed by Fine in \cite{Fine1982}.

Let $\mathscr{C}$ be the state space where generic $\epsilon$ reside and $\mathscr{C}_\mathcal{Y}$ the state space for $\mathcal{Y}$-realist states. While $\mathscr{C}$ is convex by construction, the convexity of $\mathscr{C}_\mathcal{Y}$ remains to be proved. Considering the one-parameter state $\epsilon'_\lambda = (1-\lambda)\epsilon'_1 + \lambda \epsilon'_2$, with $\lambda \in [0, 1]$ and ${\epsilon'_1, \epsilon'_2} \in \mathscr{C}_\mathcal{Y}$, the proof is complete if $\epsilon'_\lambda \in \mathscr{C}_\mathcal{Y}$. Suppose a binary random variable $\Lambda$ with assigned probabilities $(1-\lambda)$ and $\lambda$. A rule may be given such that, depending on the outcome of $\Lambda$, $\mathcal{X}$ will be measured in either $\epsilon'_1$ or $\epsilon'_2$. The probabilistic profile for this scenario is represented by $p_{\epsilon'_\lambda}(x_i)=p_{(1-\lambda)\epsilon'_1 + \lambda \epsilon'_2}(x_i)$. Due to the $\mathcal{Y}$-realism status of $\epsilon'_{1,2}$,
\begin{equation}
    p_{(1-\lambda)\epsilon'_1 + \lambda \epsilon'_2}(x_i) = p_{(1-\lambda)\Phi_\mathcal{Y}(\epsilon'_1) + \lambda \Phi_\mathcal{Y}(\epsilon'_2)}(x_i).
\end{equation}
However, the right-hand side of this equation can also represent a preparation and mixing of two ensembles of $\epsilon'_1$ and $\epsilon'_2$ with relative populations $(1-\lambda)$ and $\lambda$, respectively, with the measurement of $\mathcal{Y}$ first and $\mathcal{X}$ second from a randomly selected state from the mixed ensemble, $\epsilon'_\lambda$. That is,
\begin{equation}
    p_{(1-\lambda)\Phi_\mathcal{Y}(\epsilon'_1) + \lambda \Phi_\mathcal{Y}(\epsilon'_2)}(x_i) = p_{\Phi_\mathcal{Y}(\epsilon'_\lambda)} (x_i).
\end{equation}
This last equality, which mirrors the fact that $\Phi$ is an affine map, can be connected to the previous equations to complete the proof.

Classical mechanics gives the simplest case for realist theories. The state space $\mathscr{C}$ becomes a $2n$-dimensional phase space, where $n$ represents the dimensions of position and momentum. A state $\epsilon$ in this space is one point that characterizes a configuration of a physical system. The realism of the framework is implied by the space of $\mathcal{Y}$-realist states being equal to the total phase space, $\mathscr{C} = \mathscr{C}_\mathcal{Y}$ for every $\mathcal{Y}$. In classical statistical mechanics---that is, classical mechanics supplemented with subjective uncertainties and Liouville's equation---the diagnosis is no different. For every generalized coordinate $q_i$ and canonical momentum $\pi_j$, a joint probability distribution $\mathcal{P}(q_i,\pi_j)\,dq_id\pi_j$ can be found, which, as discussed above, validates the realism criterion \eqref{eq:criterion}.

In the mathematical framework of quantum mechanics (QM), the left and right-hand sides of Eq.~\eqref{eq:criterion} become:
\begin{align}
    p_\epsilon(x_i) &\xrightarrow{\text{QM}} \Tr [X_i \rho ], \\
    p_{\Phi_Y(\epsilon)}(x_i) &\xrightarrow{\text{\scriptsize QM}} \Tr [X_i \Phi_Y(\rho) ],
\end{align}
with $\Phi_Y$ defined as in Eq.~\eqref{eq: Phi_A}. A state $\epsilon$ is represented by a density operator $\rho$ and a physical property $\mathcal{X}$ by an observable $X=\sum_i x_i X_i$, being $X_i$ projectors corresponding to measurement outcomes $x_i$. In this context, criterion \eqref{eq:criterion} reads $\Tr [X_i \rho ] = \Tr [X_i \Phi_Y(\rho) ]$ and, because of the linearity and cyclic properties of the trace, this equation can be written as:
\begin{equation} \label{eq: qm realism}
    \Tr [X_i (\Phi_Y(\rho) - \rho) ] = 0.
\end{equation}
A sufficient condition for this equation to hold is for $\rho$ to be a $Y$-realist state according to criterion \eqref{eq:bilcrit}. The fact that this condition is also necessary is proven in Appendix A.

\section{\label{sec:level4}Quantifying irrealism}

If two generic states in a physical theory fall outside $\mathscr{C}_\mathcal{Y}$, thus violating criterion \eqref{eq:criterion} for a physical property $\mathcal{Y}$, it should be possible to quantify and compare the degrees of violation of these states. Thus, we propose two different methodologies for quantifying {\it irrealism}, the complement of realism.

\subsection{Robustness of Irrealism}

Here, we employ the usual approach of robustness under state perturbation. For a state $\epsilon$ under scrutiny, we introduce another state $\epsilon' \in \mathscr{C}$ in a convex sum weighted by a factor $\eta$, yielding a state $\epsilon_\eta=(1 - \eta) \epsilon + \eta \epsilon'$. We ask, what is the minimal $\eta$ such that $\epsilon_\eta \in \mathscr{C}_\mathcal{Y}$? Hereafter referred to as the {\it robustness of irrealism} of state $\epsilon$ regarding the quantity $\mathcal{Y}$, the optimal value of $\eta$ that answers this question is mathematically formulated as
\begin{equation} \label{eq: robus}
   \mathcal{R}_\mathcal{Y}(\epsilon)\coloneqq \min_{\epsilon'} \big\{\eta\in [0,1] \; \big| \; (1 - \eta) \epsilon + \eta \epsilon' \in  \mathscr{C}_\mathcal{Y}\big\}.
\end{equation}
We resort to the robustness $\mathcal{R}_\mathcal{Y}(\epsilon)$ because it takes advantage of the convex structure of the spaces $\mathscr{C}$ and $\mathscr{C}_\mathcal{Y}$ without relying on a particular definition of a metric for such spaces. Moreover, it satisfies desirable conditions for an irrealism quantifier. Specifically, $\mathcal{R}_\mathcal{Y}(\epsilon)\geq 0$, with equality holding if and only if $\mathcal{Y}$ is an element of reality, i.e., $\epsilon \in \mathscr{C}_\mathcal{Y}$. The requirement for $\epsilon' \in \mathscr{C}$ and not $\mathscr{C}_\mathcal{Y}$ is because if $\dim(\mathscr{C}_\mathcal{Y}) < \dim(\mathscr{C})$, then $\mathcal{R}_\mathcal{Y}(\epsilon)$ would always be zero, and this measure would not provide any relevant information about the irreality status of an $\epsilon$.

\begin{figure}[tb] 
\centering
\includegraphics[scale=0.27]{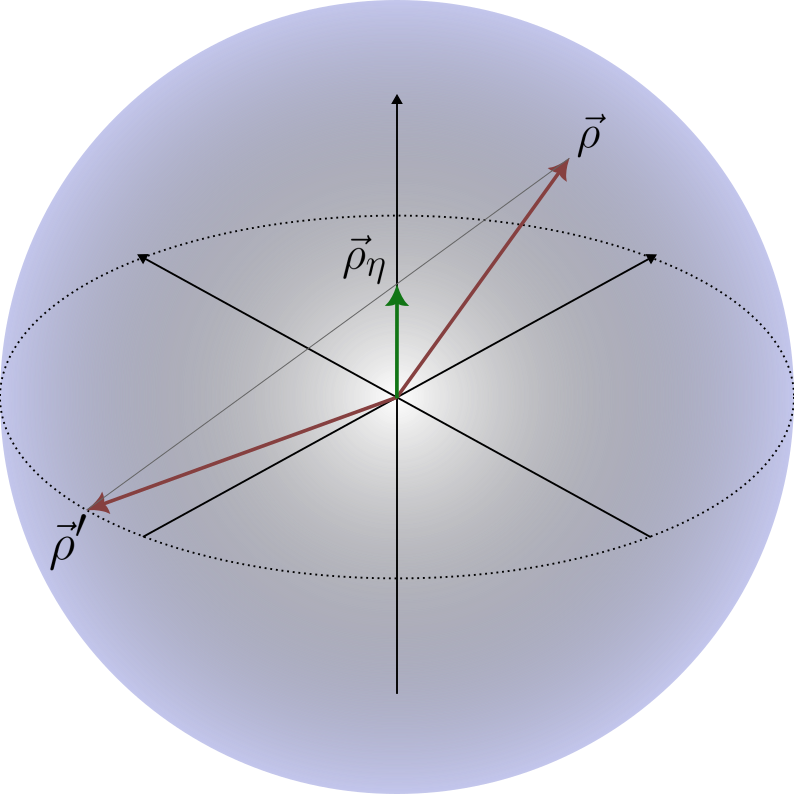} 
\caption{Bloch sphere representation for robustness of irrealism of $S_{\hat{z}}$ for a qubit. The longer vertical axis represents the set of states  $\mathscr{C}_{S_{\hat{z}}}$. The three colored arrows correspond to the vectors that parametrize the states in Eq.~\eqref{eq: robus} after the minimization of $\eta$ over $\rho'$. For $\rho=\frac{1}{2}(\mathbbm{1}+\vec{\rho})\cdot\vec{\sigma}$, given by  $\vec{\rho}=(x,y, z)\in\mathbbm{R}^3$, the robustness $\mathcal{R}_{S_{\hat{z}}}(\rho)=\bar{\eta}$ [see Eq.~\eqref{eq: geo rob}] is found for $\rho'=\frac{1}{2}(\mathbbm{1}+\vec{\rho}'\cdot\vec{\sigma})$ with $\vec{\rho}'=(-x, -y, 0)\in\mathbbm{R}^3$.}
\label{fig: bloch}
\end{figure}

One can gain insight into the physical intuition behind the concept of robustness of irrealism by envisioning a simple physical scenario. Suppose you have a given state $\epsilon$ and, at your disposal, every other state $\epsilon'$. Robustness tells you the least amount $\eta$ you could add of a state $\epsilon'$ to $\epsilon$ such that the resulting ensemble $(1 - \eta) \epsilon + \eta \epsilon'$ is a realist state for a given physical property. That is, it quantifies the minimum amount of mixing required for the irrealism of a given state to be lost.

In quantum mechanics, for a single qubit, realism as defined in Eq.~\eqref{eq:bilcrit} reduces to coherence~\cite{Bilobran2015}. The same is true for the robustness of realism. Indeed, our findings for such a case are consistent with those reported in \cite{Napoli2016}. 
%Yet, this is not the case for multipartite systems, and Eq.~\eqref{eq: robus} holds for these scenarios as well.
To show this, let us compute the robustness of irrealism of the spin observable in the $\hat{z}$ direction, $S_{\hat{z}}$, for a preparation $\rho$. For a minimization, as defined in Eq.~\eqref{eq: robus}, returning the optimal value $\bar{\eta}$, we have $\rho_{\bar{\eta}} = (1-\bar{\eta}) \rho + \bar{\eta} \rho'$, where $\rho_{\bar{\eta}}$ is a $S_{\hat{z}}$ realist state. As found in \cite{Napoli2016}, the representation of such states in the Bloch sphere reveals a geometric pattern (see Fig.~\ref{fig: bloch}): being $\rho=\frac{1}{2}(\mathbbm{1}+\vec{\rho}\cdot\vec{\sigma})$ with $\vec{\rho}=(x, y, z)\in\mathbbm{R}^3$, one finds $\rho'=\frac{1}{2}(\mathbbm{1}+\vec{\rho}'\cdot\vec{\sigma})$ with $\vec{\rho}'=(-x,-y,0)\in\mathbbm{R}^3$, where $\vec{\sigma}$ is the vector composed of Pauli matrices. Because $\bar{\eta}$ is numerically equivalent to the ratio between $|\vec{\rho}_{\bar{\eta}} - \vec{\rho}'|$, where $\vec{\rho}_{\bar{\eta}}$ is the vector representing $\rho_{\bar{\eta}}$, and $|\vec{\rho} - \vec{\rho}'|$, it is a simple matter of geometry to derive the expression
\begin{equation} \label{eq: geo rob}
    \mathcal{R}_{S_{\hat{z}}}(\rho)=\bar{\eta} = \frac{|\hat{z}\times\vec{\rho}|}{1 + |\hat{z}\times\vec{\rho}|},
\end{equation}
where $|\hat{z}\times\vec{\rho}|=r\sin{\theta}$, with $r$ being the radius and $\theta$ the polar angle defining the spherical coordinates of the vector $\vec{\rho}$. For a generic spin observable, $S_{\hat{n}}$, Eq.~\eqref{eq: geo rob} still yields the robustness of irrealism under a basis rotation, aligning $\hat{z}$ with $\hat{n}$.

\begin{figure}[t] 
\centering
\includegraphics[width=1\columnwidth]{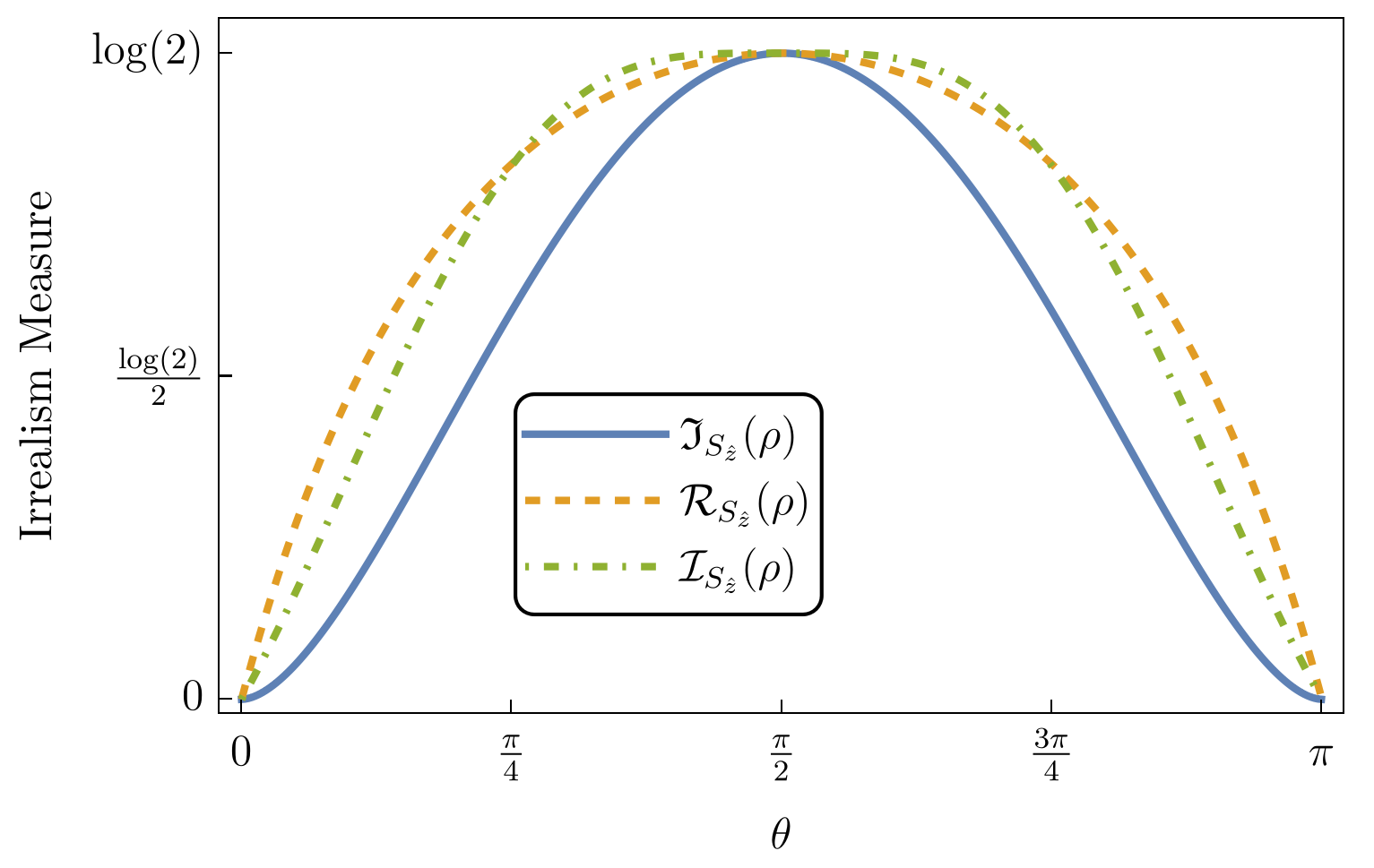} 
\caption{Plot of $\mathfrak{I}_{S_{\hat{z}}}(\rho)$, $\mathcal{R}_{S_{\hat{z}}}(\rho)$, and $\mathcal{I}_{S_{\hat{z}}}(\rho)$ for a pure qubit state with its representing polar angle $\theta$ in $[0,\pi]$.}
\label{fig: plots}
\end{figure}

Numerical comparisons between the irreality defined in Eq.~\eqref{eq: bil irr} and the robustness of irrealism are noteworthy. Figure~\ref{fig: plots} shows the results of a case study of the irrealism of $S_{\hat{z}}$ for a pure state of a qubit, where the polar angle $\theta$ in the Bloch representation varies from $0$ to $\pi$. In this computation, we normalized the robustness of irrealism so that its maximum matches the irreality's. We see that both criteria reach zero at the same points, the maximum value at the same point, and the parametric curve for both functions, displayed in Fig.~\ref{fig: param plots}, confirms a monotonic relation.

\subsection{Divergence of Realism}

\begin{figure}[t] 
\centering
\includegraphics[width=1\columnwidth]{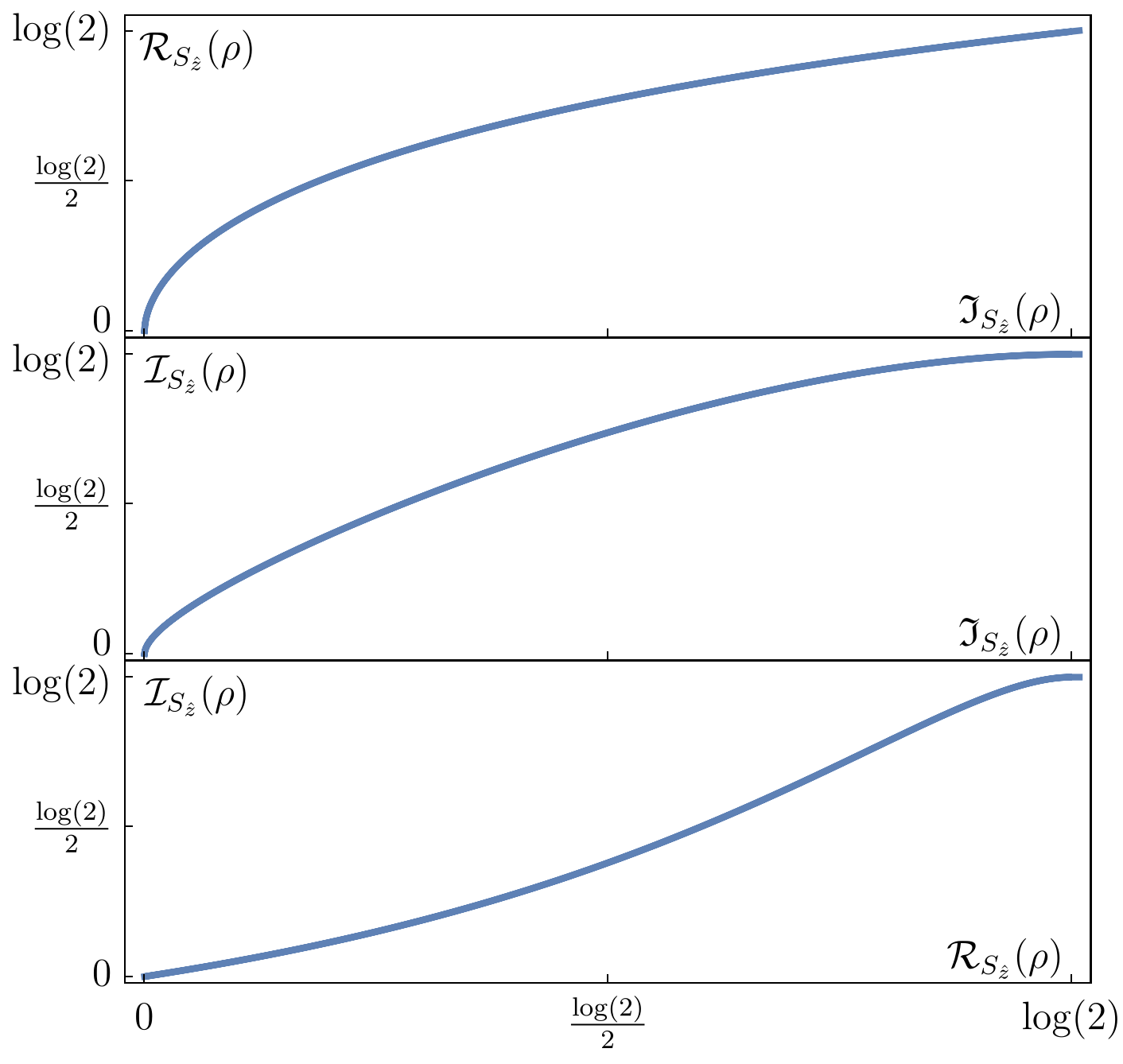} 
\caption{Three parametric plots for the same case of a pure qubit state, with the polar angle $\theta$ ranging from $0$ to $\pi$ in the Bloch sphere representation. The top plot shows $\mathfrak{I}_{S_{\hat{z}}}(\rho)$ against $\mathcal{R}_{S_{\hat{z}}}(\rho)$, the middle plot shows $\mathfrak{I}_{S_{\hat{z}}}(\rho)$ against $\mathcal{I}_{S_{\hat{z}}}(\rho)$, and the bottom plot shows $\mathcal{R}_{S_{\hat{z}}}(\rho)$ against $\mathcal{I}_{S_{\hat{z}}}(\rho)$.}
\label{fig: param plots}
\end{figure}

Definition \eqref{eq: bil irr} hints in another direction by which a theory-independent irrealism quantification may be conceived. We can simply take a divergence between the left and right-hand terms appearing in the criterion \eqref{eq:criterion}, and, for this purpose, the Kullback-Leibler divergence~\cite{Kullback1951} provides us with a standard framework. Because the criterion is defined for any physical property $\mathcal{X}$, the violation of realism is given by the property that wields the maximal divergence between the probability distributions. This leads us to the following quantifier for the irreality of $\mathcal{Y}$ given a physical state $\epsilon$:
\begin{equation} \label{eq: div irr}
    \mathcal{I_Y} (\epsilon) = \max_{\mathcal{X}}  \,\mathcal{D} \left( P^\mathcal{X}_{\epsilon} \Big\| P^\mathcal{X}_{\Phi_\mathcal{Y}(\epsilon)} \right).
\end{equation}
Here, $\mathcal{I_Y} (\epsilon)$ reads as the {\it divergence of realism} of $\mathcal{Y}$ for the state $\epsilon$, with the Kullback-Leibler divergence between probability distributions $P_\epsilon^\mathcal{X}=\{p_\epsilon(x_i)\}_{i=1}^d$ and $Q_\epsilon^\mathcal{X}=\{q_\epsilon(x_i)\}_{i=1}^d$ computed as $\mathcal{D}(P_{\epsilon}^\mathcal{X}||Q_{\epsilon}^\mathcal{X})=\sum_ip_{\epsilon}(x_i)\ln\big[p_{\epsilon}(x_i)/q_{\epsilon}(x_i)\big]$. It follows from the definition of the Kullback-Leibler divergence that $\mathcal{I_Y} (\epsilon)$ is always non-negative and vanishes {\it iff} the two probability distributions are the same, that is, if the criterion \eqref{eq:criterion} is met or, equivalently, $\epsilon \in \mathscr{C}_\mathcal{Y}$.

While we appealed to the convex geometric structure of the state spaces and to physical intuition to derive the robustness of irrealism, the divergence of realism relies solely on classical probability theory. Specifically for quantum mechanics, it is shown in Appendix B that 
\begin{equation} \label{eq: div irr quant}
    \mathcal{I}_Y (\rho) = \max_{X} \, S \big(\Phi_X (\rho) || \Phi_{X,Y} (\rho)\big),
\end{equation}
where $\Phi_{X,Y} (\rho) = \Phi_X\big(\Phi_Y (\rho)\big)$, for observables $X$ and $Y$. Equation~\eqref{eq: div irr quant}, for the case of qubits, peaks at the same point state as Eq.~\eqref{eq: bil irr}, just as the robustness of irreality. A case study employing the same states as in the preceding subsection was conducted, and the findings are illustrated in Figs.~\ref{fig: plots} and \ref{fig: param plots}. It is worthwhile noting that the apparent plateau in $\mathcal{I}_{S_{\hat{z}}} (\rho)$ in Fig.~\ref{fig: plots} is not a plateau. Instead, it is a region where the absolute value of the slope of the curve continuously approaches zero, with zero occurring only at the peak, in a gradually decreasing fashion. The monotonic relations found in Fig.~\ref{fig: param plots} among the irrealism quantifiers  $\mathfrak{I}_{S_{\hat{z}}}(\rho)$, $\mathcal{R}_{S_{\hat{z}}}(\rho)$, and $\mathcal{I}_{S_{\hat{z}}}(\rho)$ suggest that these measures may be qualitatively equivalent.

\section{\label{sec:level5}Conclusion}

Despite a rich body of literature on violations of hypotheses such as locality, causality, or even macrorealism~\cite{Legget1985} or local-friendliness~\cite{Bong2020}, realism is a notion that sparks debates over its very definition. Additionally, quantifiable notions of realism transcending the scope of quantum theory are still lacking. This work fills this gap by bringing together different criteria and aiming to provide material for the emergence of a consensus on the subject.

By only requiring the probabilities assigned to the outcomes of measurements of physical properties in a physical system, we have crafted a theory-independent realism criterion, Eq.~\eqref{eq:criterion}. It presents itself as a legitimate generalization of BA's criterion and also Fine's criterion, and it retrieves the results expected for both quantum and classical mechanics. Deviance measures for this criterion, the so-called irrealism measures, were developed in two independent ways, making use of the notion of robustness and the divergence of Kullback-Leibler. Although the robustness of irrealism [Eq.~\eqref{eq: robus}] and the divergence of realism [Eq.~\eqref{eq: div irr}] are not conceptually equivalent, neither to each other nor to BA's irreality measure, our case study highlighted a similar behavior for the three quantities. This provides evidence that both the robustness of irrealism and the divergence of realism are candidates for \textit{bona fide} irreality quantifiers.

The numerical analysis of the robustness of irrealism in a qubit system, as presented in this work, is computationally feasible. However, calculations for higher-dimensional systems require the development of sophisticated computational strategies. We expect future research efforts in this direction to yield insights. Concurrently, our ongoing research delves into the interplay between realism and context incompatibility, particularly from a theory-independent perspective. Finally, this work paves the way for a deeper understanding of the emergence of classical realism, and we expect a revisit of Wigner's friend scenarios under this light to bear fruit.

\begin{acknowledgments}

This research was financed in part by the Coordenação de Aperfeiçoamento de Pessoal de Nível Superior, Brasil (CAPES), Finance Code 001. D.M.F. thanks Eduardo Hoefel and Ana C. S. Costa for insightful conversations. R.M.A. thanks the financial support from the National Institute for Science and Technology of Quantum Information (CNPq, INCT-IQ 465469/2014-0) and the Brazilian funding agency CNPq under Grant No. 305957/2023-6.

\end{acknowledgments}

\appendix

\section{}

Our intent here is to show that Eq.~\eqref{eq: bil irr} is a necessary and sufficient condition for Eq.~\eqref{eq: qm realism}. It is an established result, shown in Ref.~\cite{Plavala2023}, that there is a duality between the state space and the effect algebra. In other words, given a state space $\mathscr{K}$, it is possible to construct the effect algebra $E(\mathscr{K})$ and vice versa. The proof shown below is a particular case of this result, relying exclusively on the formalism of quantum mechanics.

It is evident that if Eq.~\eqref{eq:bilcrit} holds, Eq.~\eqref{eq: qm realism} will hold as well. It is not immediately clear, however, that this is the only case where Eq.~\eqref{eq: qm realism} is met, and this is what we prove. To this end, we first introduce some aspects of the formalism of the generalized Bloch representation. A more complete yet concise exposition is given in sources \cite{Aerts2014} and \cite{Aerts2016}.

A basis for the linear operators acting on the state space is given by the set of matrices $\left\{\mathbbm{1}, \Lambda_1, \ldots, \Lambda_{d^2-1}\right\}$, which are the generators of the special unitary group of degree $d$, SU($d$). The $\Lambda_i$ are complex $d \times d$ self-adjoint orthogonal traceless matrices. Together with the normalization $\Tr(\Lambda_i \Lambda_j)=2 \delta_{i j}$, a generic quantum state can be expressed as
\begin{equation} \label{eq: Ap1.1}
    \rho_{\vec{r}} = \frac{1}{d} \left(\mathbbm{1} + C_d \vec{r} \cdot \vec{\Lambda}\right),
\end{equation}
where $C_d = \sqrt{d(d-1) / 2}$ and, with an orthonormal basis in $\mathbbm{R}^{d^2-1}$ denoted by $\left\{\hat{e}_i\right\}_{i=1}^{d^2-1}$, $\vec{r}=\sum_{i=1}^{d^2-1} r_i \hat{e}_i$ is a vector and $\vec{\Lambda}=\sum_{i=1}^{d^2-1} \Lambda_i \hat{e}_i$ is a vector with matrices as components. This allows for any quantum state to be represented as a real vector $\vec{r}$ in a real ball $B(\mathbbm{R}^{d^2-1})$ of dimension $d^2 - 1$. Projective operators, characterized by $\sum_j X_j = \mathbbm{1}$ and $\operatorname{Tr}\left(X_i X_j\right)=\delta_{i j}$ can be written as
\begin{equation} \label{eq: Ap1.2}
    X_i = \frac{1}{d} \left(\mathbbm{1}+C_d \vec{x}_i \cdot \vec{\Lambda}\right),
\end{equation}
where $\sum_i \vec{x}_i = \vec{0}$ and $\vec{x}_i \cdot \vec{x}_j = (\delta_{i j} d-1) /(d-1)$. Being $x_i$ the eigenvalues of a traceless observable $X$, one can write $X=\sum_ix_iX_i=\vec{x}\cdot\vec{\Lambda}$ with $\vec{x}=(C_d/d)\sum_ix_i\vec{x}_i$. With a similar construction for $Y=\vec{y}\cdot\vec{\Lambda}$, where $\vec{y}=(C_d/d)\sum_jy_j\vec{y}_j$, it can be shown that~\cite{Martins2020}
\begin{equation} \label{eq: Ap1.3}
    \Phi_Y (\rho_{\vec{r}}) = \frac{1}{d} (\mathbbm{1} + C_d \vec{u} \cdot \vec{\Lambda}), \quad \vec{u} = \frac{d-1}{d} \sum_{j=1}^d (\vec{y}_j \cdot \vec{r}) \vec{y}_j,
\end{equation}
Substituting Eqs.\eqref{eq: Ap1.1}, \eqref{eq: Ap1.2}, and \eqref{eq: Ap1.3} into Eq.\eqref{eq: qm realism}, multiplying the resulting equation by $x_i$ and summing over $i$ yields
\begin{equation}
    \Tr[(\vec{x} \cdot \vec{\Lambda}) (\vec{v} \cdot \vec{\Lambda})]=0
\end{equation}
with $\vec{v} \coloneqq \vec{r} - \vec{u}$. Now, using the identity $\Tr[(\vec{r}_1 \cdot \vec{\Lambda}) (\vec{r}_2 \cdot \vec{\Lambda})] = 2 (\vec{r}_1 \cdot \vec{r}_2)$, the last equation becomes simply
\begin{equation}
    \vec{x} \cdot \vec{v} = 0.
\end{equation}
We complete the proof by observing that this equation should hold for every possible observable $X$, characterized by $\vec{x}$, which is only possible if $\vec{v} =\vec{0}$, or, equivalently, $\vec{r} = \vec{u}$, implying Eq.~\eqref{eq:bilcrit}.

\section{}

Here, we demonstrate that, within the context of quantum mechanics, Eq.~\eqref{eq: div irr} can be reformulated as shown in Eq.~\eqref{eq: div irr quant}. To this end, we employ the relation $f(A) \ket{a} = f(a) \ket{a}$, where $f$ is a generic function and $A$ is a Hermitian operator with orthonormal eigenbasis $\ket{a}$ and eigenvalues $a$. Now, we note that $\Phi_X(\rho) = \sum_i p_\rho(x_i)X_i$ and $\Phi_{X,Y}(\rho) = \sum_i p_{\Phi_Y(\rho)}(x_i)X_i$, implying that these post-measurement states commute and therefore share the same set of eigenvectors. Then,
\begin{align}
    &S\big(\Phi_X (\rho) || \Phi_{X,Y} (\rho)\big) \notag \\
    &= \Tr\big[\Phi_X (\rho) \log \Phi_X (\rho)\big] - \Tr[\Phi_X (\rho) \log \Phi_{X,Y} (\rho)] \\
    &= \sum_i p_{\rho} (x_i) \log p_{\rho} (x_i) - \sum_i p_{\rho} (x_i) \log p_{\Phi_Y (\rho)} (x_i) \\
    &= \mathcal{D} \left( P^{X_i}_{\rho} \Big\| P^{X_i}_{\Phi_Y(\rho)} \right).
\end{align}
Since the argument inside the maximization function is equal for both expressions, so is its result. This proof was originally presented in Ref.~\cite{Storrer2023}.

\bibliography{apssamp}% Produces the bibliography via BibTeX.

\end{document}